# A Systematic Survey of the Gemini Principles for Digital Twin Ontologies

An examination of how effectively ontologies align with the Gemini Principles across time and industry sectors

James M. Tooth

Department of Science, Technology, Engineering and Public Policy, University College London, London, United Kingdom

james.tooth.21@ucl.ac.uk

Nilufer Tuptuk

Department of Security and Crime Science, University College London, London, United Kingdom

n.tuptuk@ucl.ac.uk

Jeremy D. M. Watson CBE

Department of Science, Technology, Engineering and Public Policy, University College London, London, United Kingdom

jeremy.watson@ucl.ac.uk

Ontologies are widely used for achieving interoperable Digital Twins (DTws), yet competing DTw definitions compound interoperability issues. Semantically linking these differing twins is feasible through ontologies and Cognitive Digital Twins (CDTws). However, it is often unclear how ontology use bolsters broader DTw advancements. This article presents a systematic survey following the PRISMA method, to explore the potential of ontologies to support DTws to meet the Centre for Digital Built Britain's Gemini Principles and aims to link progress in ontologies to this framework.

The Gemini Principles focus on common DTw requirements, considering: **Purpose** for 1) *Public Good*, 2) *Value Creation*, and 3) *Insight*; **Trustworthiness** with sufficient 4) *Security*, 5) *Openness*, and 6) *Quality*; and appropriate **Functionality** of 7) *Federation*, 8) *Curation*, and 9) *Evolution*. This systematic literature review examines the role of ontologies in facilitating each principle. Existing research uses ontologies to solve DTw challenges within these principles, particularly by connecting DTws, optimising decision-making, and reasoning governance policies. Furthermore, analysing the sectoral distribution of literature found that research encompassing the crossover of ontologies, DTws and the Gemini Principles is emerging, and that most innovation is predominantly within manufacturing and built environment sectors. Critical gaps for researchers, industry practitioners, and policymakers are subsequently identified.

# 1 INTRODUCTION

Digital Twins (DTws) are a widespread solution that are beginning to be explored for real-time monitoring, simulation, and analysis of physical processes across a wide range of distinctive applications. Interoperability, allowing heterogeneous information for different systems, devices, or applications to work together effectively [1] is required for systems-level federation; yet, design differences across different DTw applications can hinder interoperability between DTws. One solution could be the application of ontological methods, which are capable of providing a common representation of data between different systems. However, embedding ontologies into DTws poses a question, which is explored as the contribution of this paper; **how effectively do applications of ontologies align with existing Digital Twin frameworks, and how does this compare across different sectoral applications?** Answering this question could support evaluation of ontologies as part of broader DTw evaluation.

DTws originated concurrently from two domains 1) Product Lifecycle Management (PLM) [2] in Industry 4.0 [3], and 2) Building Information Modelling (BIM) from the built environment [4]. The first fully formed concept of DTws was introduced in Michael Grieves' 2003 definition, which consists of *physical products* (the assets a user wishes to digitalise), *virtual products* (the digitised models of the assets) and *connections* (the sensors and tools which link assets to models) [5] and is acknowledged as an effective baseline definition by many sources [3], [4], [6], [7], [8], [9], [10], [11], [12], [13], [14], [15], [16], [17], [18], [19], [20], [21], [22], [23], [24], [25], [26], [27], [28], [29], [30], [31], [32], [33], [34], [35], [36], [37], [38], [39]. However, others have suggested limitations of this definition include an over simplification of the complexity of the structure of DTws [40], a lack of focus on accurately mirroring the physical asset in real time [41] and little consideration of evolving systems or applications outside of the manufacturing domain [2]. Consequently, other definitions of DTws have arisen (which are investigated in section 2.1.1).

Such a range of definitions is understandable for DTws within literature, considering their diverse applications. These applications include a wide variety of use cases including: monitoring assets to detect, diagnose, analyse and manage changes, anomalies, faults or conflicts in data or parameters; verifying instructions, controlling assets and making automated decisions; predicting and evaluating hypothetical scenarios to optimise asset or component performance [3], [9], [11], [16], [40], [42], [43], [44]. Furthermore, whilst many innovations such as Industry 4.0, cyber physical systems, BIM and UAVs [4], [6], [16], [45], [46], [47] are instanced as individual technologies, DTws could be most valuable as a common platform synergising with these technologies, data and assets [40], [48]. Whilst DTws seem promising for future developments, interoperability challenges present themselves as DTws and their capabilities are yet to be fully defined and agreed upon [40]; the broader concept of a DTw as a physical system providing and using data does seem widely agreed between sectors, yet literature suggests inconsistency arises as most DTw research is conducted inside sectors, rather than methodically working with all data across sectors [49]. Since different users and vendors may prioritise different capabilities, granularities and connections [50], this lack of consensus can lead to conflicting expectations, requirements and DTw designs [51] (as explored in section 2.1.2).

To address this complexity, the adoption of ontologies is being considered as a potential solution. This approach uses metadata to map out the relationships between different data sources and assets, in order to create common reference information linking disparate DTws [8], [9], [37], [43], [52], [53], [54] (and are explained in section 2.2.1). Furthermore, synergising data and knowledge by integrating ontologies (and Machine Learning) into DTws forms a novel innovation termed Cognitive Digital Twins (CDTws) [12]. Some literature suggests CDTws may enhance the biggest benefits of a DTw, to more accurately represents the physical object [18] and to respond and optimise the supply chain with real-time data [13], [16]. Given the breadth of ML, this paper focusses on the ontological aspects of CDTws (section 2.2.2).

With such a broad environment of opportunities and applications for DTws, a one-size-fits-all approach to ontologies may hence be inappropriate, and different approaches may be required for various DTw systems (as detailed in section 2.3.1). As such, a framework approach may be beneficial to identifying the most favourable uses of a DTw in any particular circumstances. One possible existing framework for evaluating the effectiveness of DTws is the Gemini Principles, developed by the Centre for Digital Built Britain (CDBB) in 2018, to help businesses develop DTws that can be integrated into a National Digital Twin (a national ecosystem spanning connected DTws) [55]. This framework encompasses nine principles: 1) *public good*, 2) *value*, 3) *insight*, 4) *security*, 5) *openness*, 6) *quality*, 7) *federation*, 8) *curation*, and 9) *evolution*. Detailed explanations of each principle can be found in section 2.3.2.

This section introduces the topics of this paper, followed by an exploration of the rationale (1.1) and justification of the scope (1.2). Section 2 presents the literature surrounding the issues raised above, providing in-depth discussions of topics such as "*what divergences exist between different Digital Twins*?" (2.1), "*how can ontologies augment Digital Twins*?" (2.2), and "*how can Digital Twins and ontologies be evaluated*?" (2.3). Section 3 explains the methodology underpinning the systematic literature review (3.1), evaluates its suitability (3.2) and reflects on its findings (3.3). Section 4 examines how ontologies can contribute to each of the nine Gemini Principles across the three Lenses of *purpose* (4.1), *trust* (4.2) and *function* (4.3), then summarises these considerations (4.4). Finally, Section 5 outlines the gaps in identified research, policy, and industry (5.1), before concluding and summarising the work (5.2). The Appendix categorises the systematically reviewed literature by sector and date.

## 1.1 What is the rationale behind this research?

This work investigates the potential role and applicability of ontologies to support DTws in addressing each Gemini Principle in turn; doing so could align progress in ontologies with wider DTw innovation and prioritise areas for further ontology research. Additionally, this paper collates existing research about how ontologies have been considered alongside one or more Gemini Principle across sectors and time. This aims to both assess existing levels of unconscious alignment between ontology-driven DTws with the Gemini Principles, and to highlight existing areas of focus or inactivity. Consequently, this study seeks to emphasise the significance of this research direction, and to map insights and gaps for future research, policy development, and industry actions to implement ontologies.

This focus seems not to have been directly addressed by existing reviews. D'Amico covered Key Performance Indicators (KPIs) for CDTws in a maintenance context [6], whilst Ng looked at the context of digital fabrication [56]. Other systematic reviews have worked within specific sectors of supply chains [57], manufacturing [58], [59], and smart cities [51], [60], [61], [62], [63], each considering ontologies as part of a suite of solutions. This review takes a more cross-sectoral approach, linking to a framework to allow evaluation of different lenses with a view to supporting effective and connected implementation. Such research may become significant with the increased adoption of DTws. These issues may be particularly relevant now, whilst DTws are still novel and therefore before conflicting and incompatible DTw approaches become established.

## 1.2 What is the scope of this review?

This article presents a Systematic Literature Review (SLR) to assess the breadth of how ontologies facilitate application of the Gemini Principles in DTws, and understand research spread across sectors and time. The paper identifies research gaps and offers recommendations to industry and policymakers. The following research questions are addressed:

1. **How could ontologies help Digital Twins in complying with the Gemini Principles?** To establish and consolidate the current state of progress and to identify areas of agreement and debate, so that future work with ontologies can be more readily compared against the broader DTw landscape.
2. **How are publications relating to ontologies and one or more Gemini Principles distributed across sectors and over time?** To understand existing or future evolving research areas and to identify synergy or progress across sectors, so that future work can identify and action cross-sectoral gaps and learnings.
3. **What are the research, policy, and industrial gaps exist that could be addressed by ontologies to enhance Digital Twin compliance with the Gemini Principles?** To identify key areas where ontologies can improve DTw design and implementations, helping relevant stakeholders to prioritise future work.

## 2 RESEARCH CONTEXT

The broader research context covered in this review is shown in Figure 1. Fundamentally, literature is divided over how narrow DTw definitions should be, in regard to the necessity of particular requirements. On one side, some literature suggests that universal, distinct definitions and conceptualisation can clarify different terms [64], ease data sharing [55] and aid widespread understanding of what a DTw is [65]. On the other hand, certain literature suggests flexible definitions prevent unnecessary constraints and instead allow for focus on outcome and purposes [66] as well as cultivating new perspectives [48]. However, yet another literature viewpoint sees 'DTw' as a marketing phrase products are branded with [4], surrounded by hype and trendy, overused buzzwords that are in reality umbrellas for various innovations [10] which may already exist under different names [6], [14], [20].

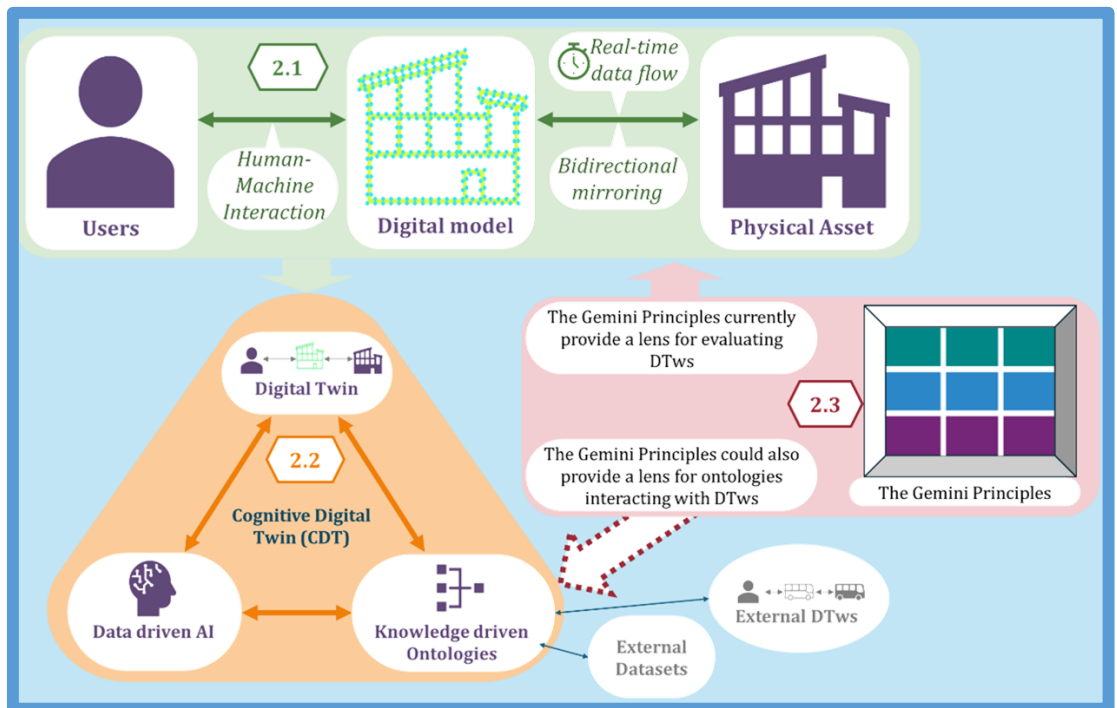

*Figure 1: Summary of the research context of this review: definitions of a DTw are explored in 2.1, CDTws and ontologies in 2.2 and the applicability of the Gemini Principles in 2.3*

Reconciling conflicting DTw perspectives infers that some distinctions may be unnecessary and undesirable [10] but that terminology should be defined for genuine clarification [10], with the phrase 'DTw' cutting across sectors and reducing silos (where unconnected assets and isolated datasets prevent sharing or collaboration) by replacing sector specific phrases such as BIM or PLM [6].

**2.1 What divergences exist between different Digital Twins?**

Additional definitions(s) of DTws may arise or evolve as the topic is viewed from different perspectives [6], [16]. Thus, futureproofed DTw networks may need capabilities to collaborate with novel DTws developed under evolving definitions. Yet, meaning may be inconsistent, varying and contradictory even with present definitions [67], [68] which were created by industry to meet sectoral needs [69] or business requirements [43], using differing protocols and standards [31]. An expanded version of the Grieves definition discussed in this subsection is shown in Figure 2.

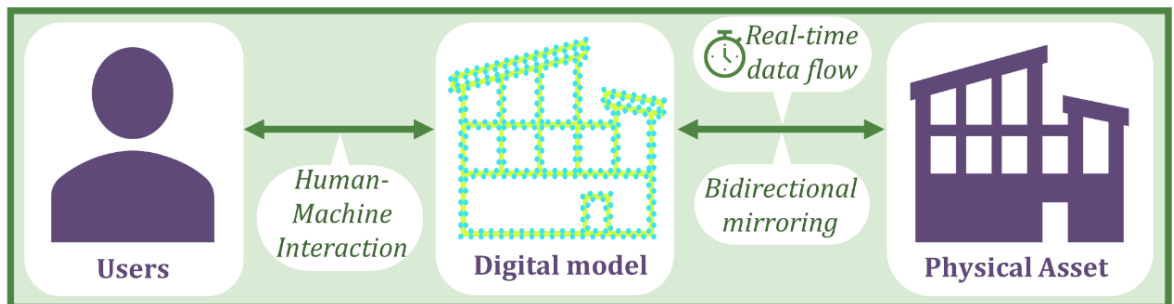

*Figure 2: Expanded Grieves definition of a Digital Twin*

*2.1.1 Contextualising Digital Twin requirements – What are the definitions of a Digital Twin?*

Literature seems to have largely developed in accordance with Grieves' definition with some developments converging on common requirements, and others diverging over new use case applications. These include AMRC's definition of "*a live digital coupling of the state of a physical asset or process to a virtual representation with a functional output*" [70], IBM's definition of "*a virtual representation of an object or system that spans its lifecycle, is updated from real-time data, and uses simulation, machine learning and reasoning to help decision making*" [71], and CDBB's definition as "*a realistic digital representation of assets, processes or systems in the built or natural environment*" such that data is connected to the physical object with real-time data from sensors [55]. One study found there are many different definitions, themselves formed from multiple definitions, and suggests there is no agreement on DTw composition, despite large quantities of literature [50]. Yet, across these varying DTw definitions are several recurring requirements:

1. **Bidirectional mirroring**: Literature advises that physical assets and virtual models communicate in both directions [2], [13], [18], [40], [42], [52] to faithfully echo each other as much as possible [4], [53], [72], [73], [74]. Whilst technically complex, mirroring may provide critical benefits for DTws [40] by reflecting relationships between assets and models via coded physical laws, accurate object positioning and a time domain representation [2]. Such mirroring can emulate a whole asset across its full lifecycle [43], [71] and extend between assets, if inter-DTw relationships are equivalent to those between the real world assets they represent [64] including hardware and software dependencies [10]. Conversely, there is division over what level of connection between assets and DTws is actually needed [50]; for instance, partial DTws called Digital Shadows primarily engage from the physical asset to digital [24], and could allow interactions from digital to physical to be gradually included as needed [21].

2. **Real-time data**: Up-to-date DTws expediently depicting real-world assets with real-time data [3], [6], [71], [73], [75], [76], [77] may be key differentiators of DTws from otherwise static asset models [6] such as BIM or PLM. Yet, real-time validation, interpretation and use can be challenging [4], particularly for heterogenous data of mixed origins [37], particularly given the bandwidth needed in sensor channels to guarantee faithful representation in the DTw of the real world instance at appropriate frequency. Nevertheless, real-time connections may not always be needed for the whole asset lifecycle, depending on how static (stays the same) or dynamic (is updated) data is [73], as well as the operation and use case [14]. Indeed, AMRC consider live to mean sufficiently low delay to enable appropriately real-time decision-making [70].
3. **Human-Machine Interaction (HMI)**: DTws are sociotechnical systems, that can communicate with both users and other digital systems, including DTws [51], [78], supporting HMI through human and machine readable outputs, configuration or control settings, and user interfaces [50]. Literature suggests fusing human resourcefulness (for unique or complex tasks) with machine efficacy (for repetitive or clear tasks) may: improve decision-making, asset efficiency and quality; reduce failures and make systems respond more proactively to incidents for improved safety; support user training, and remote work and operation; and allow collaboration across team or sectoral boundaries [3], [24], [41], [74], [78], [79], [80], [81]. Furthermore, greater efficiency gains may arise as DTws, physical assets and humans are more linked [82].

Crucially for this work, existing literature indicates that ontologies and Cognitive Digital Twins (CDTws) (which will be covered later), can meet all three of these requirements [13], [25], [52], [79].

*2.1.2 Navigating Digital Twin ambiguity – What is the variation between Digital Twins?*

Variation between DTws continues in how assets are linked. Literature suggests most DTws represent a single asset (one-to-one), and allow isolated modelling of use cases at specific lifecycle stages, rather than representing multiple assets (one-to-many) in a system of systems [2], [31], [44]. Unsurprisingly, distinct sectoral use cases, scales and lifecycle stages can hamper cross-sector interoperability and knowledge sharing as a result of discrepancies in: DTw purposes, model priorities, creation methods and data use; operational challenges in meeting differing time, sector and user needs; and a dissimilarly structured mix of open-source or commercial solutions for different components [9], [10], [20], [25], [31], [44], [51], [64], [66], [83].

Indeed, literature suggests DTws may be better suited for asset-heavy industries and complex, multistakeholder projects, rather than existing single function models (such as BIM or PLM), if they can 1) serve multiple purposes and functions, 2) process, learn and make data-led decisions from data collected by multiple stakeholders, 3) dynamically adapt data scales, properties shown, levels of details and complexity, to vary with asset lifecycle stage or user background [16], [34], [84], [85]. However, linking DTws designed with different purposes in mind could cause issues in: undefined model boundaries; terminology overlap or inconsistency; and challenges in extracting heterogenous data at high volume and rate from silos across sectors, stakeholders or lifecycle stages [10], [31], [50], [59], [75], [81].

In the resulting DTw landscape, concepts appear aligned, whilst specific properties, background, language and frameworks differ, suggesting a need to identify common features and nomenclature for DTws across sectors [2].

## 2.2 How could ontologies augment Digital Twins?

Whilst thematically similar, these definitions and requirements appear to subtly differ, with potential interoperability issues in connecting DTws developed from different definitions. Working to both standardise definitions and support nuanced differences may seem paradoxical at first, yet literature demonstrates this can be achievable using ontologies. The relationship between ontologies and CDTws discussed in this subsection is depicted in Figure 3.

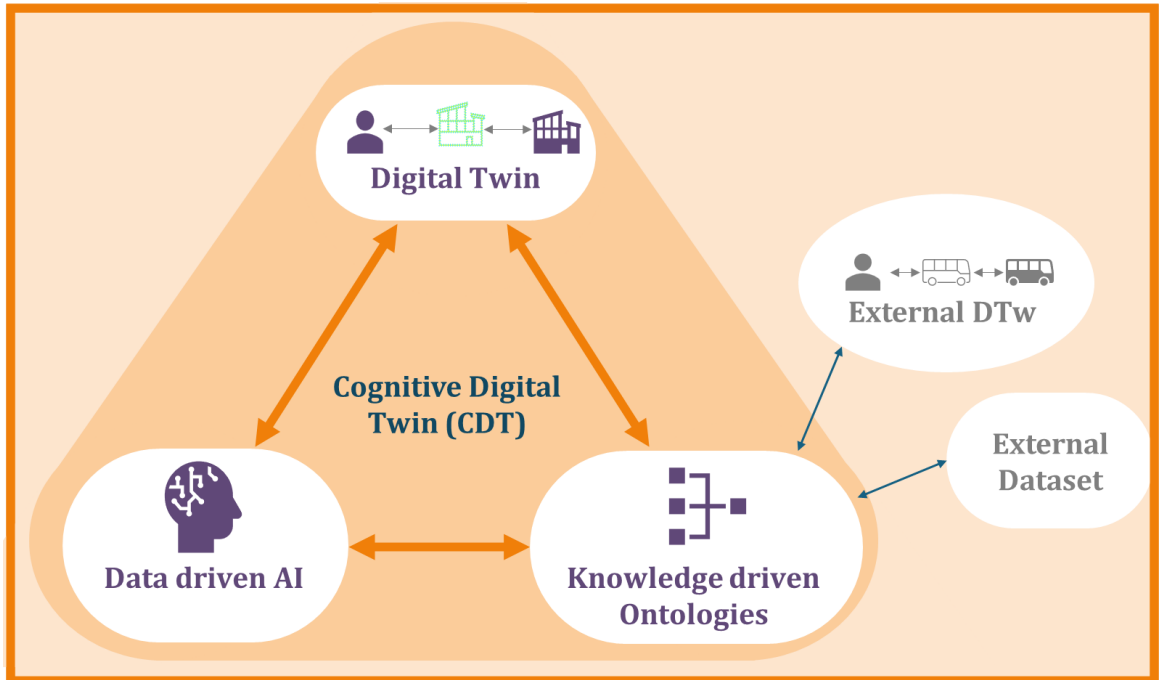

*Figure 3 Cognitive Digital Twins and ontologies*

### 2.2.1 Structuring Digital Twin data – What are Ontologies?

Philosophically, ontology is a branch of metaphysics which focusses on defining what something is and categorising objects or concepts to describe characteristics [86]. Similarly, data ontologies (simply ontologies from hereon) collate metadata (data which describes what data is included) to create a shared understanding of information about an asset. Relationships and hierarchies between metadata properties are mapped out to model how assets and data sources relate across domains or lifecycle stages. Rules can then be created or deduced to extract particular information, provide context, and geographically place information [8], [9], [37], [43], [52], [53], [54].

Through these capabilities, ontologies can deduce the consequences of how assets interact (intuitiveness), reason optimal decisions to take (cognition), and allow users to easily add new rules or relationships (extensibility). Ontologies also translate data to a common format for simplified analysis, easier information reuse, and data standardisation which can in turn support interoperable or integrated DTws, leading to sectoral knowledge being shareable with generic DTw software [4], [21], [31], [87], [88], [89]. Resulting semantic (based on words or meaning) descriptions are both human and machine readable, allowing human stakeholders to access, interpret and interact with data [90]. These descriptions can function across DTw lifecycle stages, meaning ontologies can formalise knowledge as assets are designed, then support interoperability across systems once assets are operational [52]. Semantic definitions, constraints and rules used should be as flexible, generalisable and standardised as possible to support interoperability [15] and might use the Web Ontology Language (OWL), the widely used language for ontologies [37] commonly employed with the Protégé software [52].

Beyond individual ontologies for particular systems, data from cross-sectoral DTws can be integrated through top-level ontologies (an ontology of ontologies). Whilst specific ontologies can structure information from a specific sector, challenges can arise in managing knowledge from across sectors [91], as concepts may have different meanings and objectives depending on the context [92] and the most appropriate data structure may vary [4]. As a solution, ontologies for specific sectors or lifecycle stages can be integrated through a cross-sectoral top-level ontology [6], [8], improving interoperability and reusability [91], and formalising how the relevant assets, DTws and sectors are related [12], [31], [65]. Top-level ontologies can also enable the creation of DTw metamodels, which abstractly present only high-level information. Such metamodels may offer more stability than linking many individual data models, and can show adaptable levels of detail depending on user needs, or vary the level of functionality for particular lifecycle stages or dimensions [43]. If the underlying ontologies are flexible and configurable, this could foster interoperable, reusable and resilient DTws eliminating the need to for users to have programming knowledge [19], [52].

*2.2.2 Bridging Digital Twins and Ontologies – What are Cognitive Digital Twins?*

Current DTw decision-making can be augmented with approaches that are 1) *data-driven* through AI including Machine Learning (ML) (non-traceable algorithms that learn from data), or 2) *knowledge-driven* through ontologies. Cognitive Digital Twin (CDTw) combine approaches so a integrates both principles of ontologies and ML into DTw systems, to utilise benefits and minimise limitations of each [28]. CDTws have no singular definition but are DTw-based, capable of combining pre-defined rules to reason insights through ontologies, and can autonomously integrate data across sectors and over the full or partial asset lifecycles [12], [31], [91]. The absence of a concrete description is perhaps unsurprising as CDTws are much less mature than DTws and are largely theoretical, with few successful examples [31], meaning the full capabilities of interacting CDTw networks may still be uncertain [79]. Regardless, literature proposes that in addition to the capabilities of DTws, a CDTw also needs to autonomously: 1) semantically reason (deduce rules and responses by combining existing information) to learn or predict outcomes, for data-driven decisions that improve asset resilience or sustainability; 2), transparently control and adjust the asset in real-time; 3) accurately interact with humans, assets and other DTws in a responsive way that enables commonplace issues to self-resolve, saving human cognition for complex issues; 4) continually link assets and models from other CDTws across network and sector boundaries; 5) understand the context of historic data and detect anomalies; 6) model, calibrate and optimise assets throughout the full lifecycle. As such, CDTws could become more effective the longer they are deployed in a system [2], [3], [13], [31], [41], [55], [79], [82], [93].

Literature suggests that fully standardised, interactive CDTws may perform more efficiently and be more useable [16]. However, business models to be considered in industrial deployments, and literature notes that financial and resource demands for implementing CDTws appear to be greater than for simpler DTws, due to more complex architecture which comprises multiple lifecycle stages of multiple components [31]. As such, some literature suggests CDTws should only be used as needed by the situation or stakeholders, due to the greater risk, cost and time involved [31]. CDTws could particularly benefit data reuse in complex production systems with multiple production styles and various lifecycle stages or stakeholder components that may require qualities such as flexibility, resilience, responsiveness, decision-making or autonomy [31], [34]. Given the range of use cases and resource demands surrounding DTws, that ontologies or CDTws face, it may therefore be prudent to identify what level of implementation is most advantageous.

**2.3 How can Digital Twin and ontology adoption be evaluated?**

Within the above sections, it is clear that design decisions can produce variants of DTws with divergent capabilities to address differing needs, meaning a one-size-fits-all approach is not appropriate. A framework-based approach may be beneficial to make sense of this complexity and to methodically inform users about considerations that could determine the optimum scale and extent of implementation to be adopted, as depicted in Figure 4.

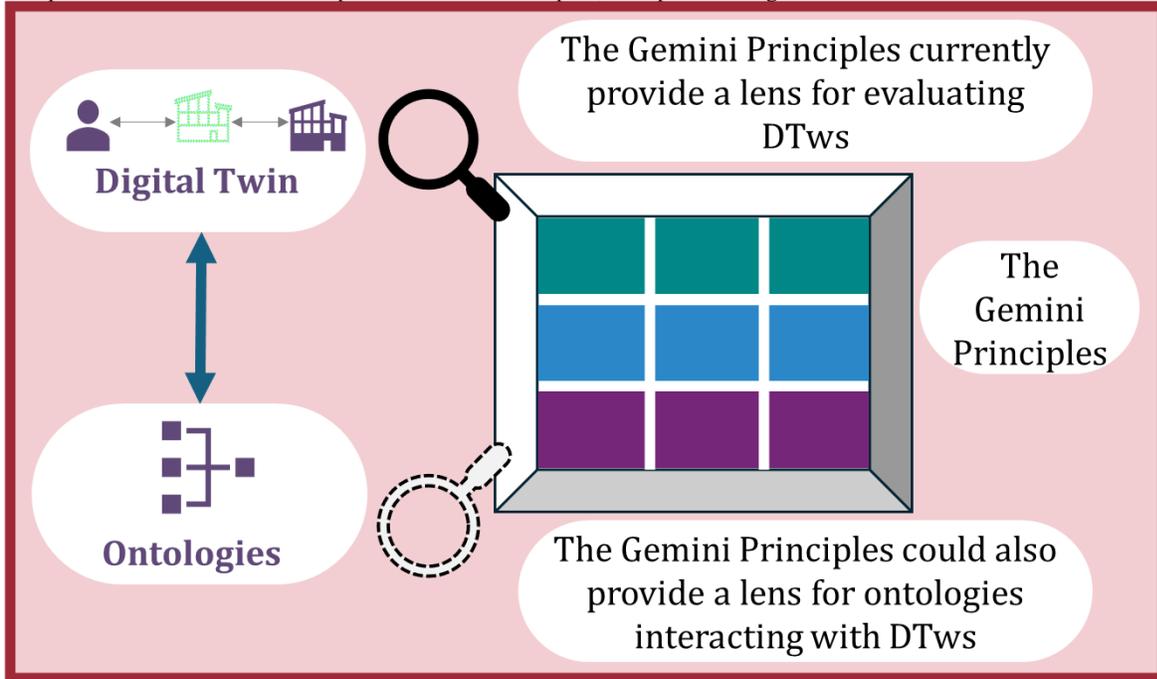

*Figure 4 How the Gemini Principles could be used to evaluate DTws and ontologies*

*2.3.1 Taking a pragmatic approach – What is necessary for a useful Digital Twin and Ontology?*

DTw research is split into conceptual research (considering what are ideal, theoretical *DTws* for the long-term), or realisation research (crafting operational DTws using existing tools to solve a particular engineering problem for near-term benefits) [23]. However, there are issues with solely pursuing either approach. Following a conceptual approach, a huge DTw could eventually link all DTws of all use cases to enable interactions and synergies between these [14], twinning assets and resource flows nationwide (termed a National Digital Twin) [55], [66], [94]. However, literature suggests that with current capabilities, complete DTws realistically will be either left incomplete, isolated and inapt, or become too unwieldly, and computationally and financially expensive [88], [92], [95]. Conversely, following a realisation path results in DTws that are fragmented mappings of data for a particular purpose, which are useful today but may not be scalable or applied beyond this scope [95]. At the extreme, partial representations with data, metadata and models to solve a specific problem and context are termed Digital Shadows [10], [59]. There are clearly advantages and disadvantages to both viewpoints.

Yet, pragmatic, intermediately-positioned DTw research focused on futureproofing realisation DTws whilst implementing conceptual DTws [31] is proposed by some literature, but not broadly defined [10], and could combine cost effective DTws and rich semantic information [95], by ontologically modelling only relevant relationships [88]. Shifting from perfect DTws towards incomplete yet practical DTws may be advantageous, given the complexity of integrating DTws and the range of DTw definitions [10]. A similar logic could apply to how ontologies are adopted. A more sophisticated, global ontology could integrate future DTws [72] but may not be immediately feasible or necessary; instead flexible ontologies that continually evolve to align with new requirements may be preferable [92]. Delineating this pragmatic approach to useable yet futureproofed DTws and ontologies is thus a research gap; this work proposes that using a framework led approach could be beneficial for this middle ground, to guide inclusion or exclusion of tools and to evaluate impacts and capabilities.

*2.3.2 Evaluating ontologies in a DTw framework – What are the Gemini Principles?*

One such framework for planning tools and impacts is the Gemini Principles, which seeks to coordinate stakeholders around a common implementation of DTws. Advantages of this framework are a focus on implementation from the individual DTw scale all the way to creating a National Digital Twin, a preference for connecting different assets as appropriate, and an aspiration to build a consensus between policymakers, asset owners and operators, investors and industry [96]. Figure 5 shows nine Gemini Principles centred around three lenses of Purpose, Trustworthiness and effective Function [55]:

| | | | |
|---|---|---|---|
| **A \| Purpose** Must have clear purpose | **1 \| Public good** Must be used to deliver genuine public benefit in perpetuity | **2 \| Value creation** Must enable value creation and performance improvement | **3 \| Insight** Must provide determinable insight into the built environment |
| **B \| Trust** Must be trustworthy | **4 \| Security** Must enable security and be secure itself | **5 \| Openness** Must be as open as possible | **6 \| Quality** Must be built on data of an appropriate quality |
| **C \| Function** Must function effectively | **7 \| Federation** Must be based on a standard connected environment | **8 \| Curation** Must have clear ownership, governance, and regulation | **9 \| Evolution** Must be able to adapt as technology and society evolve |

*Figure 5 Gemini Principles, adapted from CDBB [55]*

A. **The Purpose of DTws should be for 1) Public Good, 2) Value Creation, and 3) Insight.** Literature suggests it should be clear why a DTw will be created, how it will be created and how it will be used [10] to determine what data is collected and used [78]. *Public Good* can be considered to cover broader societal benefits beyond the immediate remit of the DTw, *Value Creation* is provision of economic benefits or performance enhancement exceeding the costs of implementation, *Insight* is new information learnt about the asset, sector or users.

B. **DTws must be Trustworthy with sufficient 4) Security, 5) Openness, and 6) Quality**. DTw business models may benefit from trusting and collaborative relationships rather than wholly transactional, competitive use [66], [97].To facilitate this trust, DTws and related models may need to become more trustworthy, and overcome liability gaps [93]. *Security* is protection of access and prevention of misuse. *Openness* is transparency in structure and sharing of data and code. *Quality* is about detecting and enforcing data accuracy and precision.

C. **DTws must have Functions of 7) Federation, 8) Curation, and 9) Evolution**. Whilst the Principles describe requirements, these are intentionally open-ended to support user innovation [55]. *Federation* is DTws working in a standard data environment that facilitates interoperability and connectivity. *Curation* ensures ownership and clear liability *Evolution* allows DTws to be extensible and adaptable to new innovations and approaches.

## 3 GENERAL METHODOLOGY

This study consists of a Systematic Literature Review (SLR). An SLR was deemed the most appropriate methodology to collate existing work in a fair, wide-reaching way as an overview of current state of the art and to provide evidence of Gemini Principles in action, even in cases where works may vary in approaches or do not consciously consider these principles. Furthermore, the reproducibility of an SLR is also important, especially in light of the fast-changing pace of this research area.

### 3.1 Systematic Literature Review description

This SLR followed the *Preferred Reporting Items of Systematic reviews and Meta-Analyses* (PRISMA) method shown in Figure 6. This method originates in healthcare, yet has been applied to SLRs across sectors [98] and consists of 3 stages:

1. **Identification** – *systematic literature search, alongside some relevant reports. In this stage duplicates are identified and removed.* In this review, literature was systematically identified from 5 major databases (ACM, IEEE, ProQuest, Scopus, Web of Science). Articles were filtered to peer-reviewed articles only. To ensure articles were directly relevant (rather than simply making passing mention), keywords needed to be in the article title, abstract or keywords. Before screening, duplicate reports were removed. In total, 416 articles were identified, with 158 duplicates removed. Keyword terms searched were:

    **"digital twin"** *to ensure the literature explicitly references Digital Twins.*

    *AND*

    **semantic OR ontolog\* OR cognitive** *to ensure an ontologically relevant aspect is mentioned, with inclusion of cognitive, to account for CDTws.*

    *AND*

    **public good OR valu\* OR insight\* OR (secur OR resilien\*) OR (open\* OR transparen\*) OR quality OR (federa\* OR interoper\* OR interrelat\* OR interconnect\* OR interact\* OR interfac\* OR integrat\* OR communicat\*) OR curat\* OR evol\*** *to ensure at least one term relating to a Gemini Principle is mentioned. Common synonyms or related themes for these were included as alternate search terms, given that global acceptance of the Gemini Principles has*

> *yet to be achieved. These were iteratively added into the search as they were encountered in literature.*

Other key reports relating to DTw definitions or the Gemini Principles and the systematic study method were also included separately, so these concepts could be explored (yet were not included in the domain evaluation later). These were attained through Google Searches for "*CDBB*", "*Grieves Definition*", "*Digital Twin definition*", "*PRISMA method systematic literature review*" and "*interrater reliability method systematic literature review*".

2. **Screening** – *literature shortlist with unsuitable work removed following inclusion or exclusion criteria (based on the abstract and title, then later full text) or that could not be found.* Within this work, articles were screened for relevance by examining the title and abstract, to ensure they related to software ontologies and DTws. All records were successfully retrieved. Next, report eligibility was assessed through inclusion criteria of the paper content making notable mention of 1) DTws, 2) ontologies/CDTws or the semantic web and 3) a concept similar to a Gemini Principle. Exclusion criteria are results not being a peer reviewed article. As a result, 8 articles were screened from an irrelevant abstract and 118 removed for not meeting the inclusion criteria.

3. **Included** – *work is collated and content analysed.* Ultimately, 130 articles were systematically obtained in this review. Separately all 13 reports identified were used. This resulted in 143 pieces of literature to be reviewed.

However, PRISMA does not account for the subjectivity in how papers are judged to meet inclusion and exclusion criteria; evaluators can have different views over how restrictive or all-encompassing to be, diminishing the replicability of work. A fourth stage of 'Inter rating' was added to evaluate the validity of the exclusion and inclusion criteria. This approach can improve reliability and lessen subjectivity in how judgements are made [99], [100].

4. **Inter rating** – *a number of the total articles are randomly selected and evaluated by a second evaluator. This number should be proportional to the number of articles and level of criteria to allow sufficient analysis without the coder overthinking. High agreement over article inclusion and exclusion indicates robust criteria.* In this work a second evaluator reviewed 25% of articles (65/260) based on full paper content and determined which papers to include or exclude, based on the above criteria. This achieved 83% agreement with the first evaluator.

Upon obtaining all of the literature, this study sought to classify articles by domain application.

5. **Context classification**: *articles are then classified by domain and year to determine the spread of literature, with the context of each work briefly described.* The domain contexts of systematically reviewed literature are shown in Figure 8 in the Appendix. Whilst articles tackled distinctive problems within varying applications, these have been categorised under the umbrella sectors of *manufacturing*, *built environment*, *other applications* and *theoretical*.

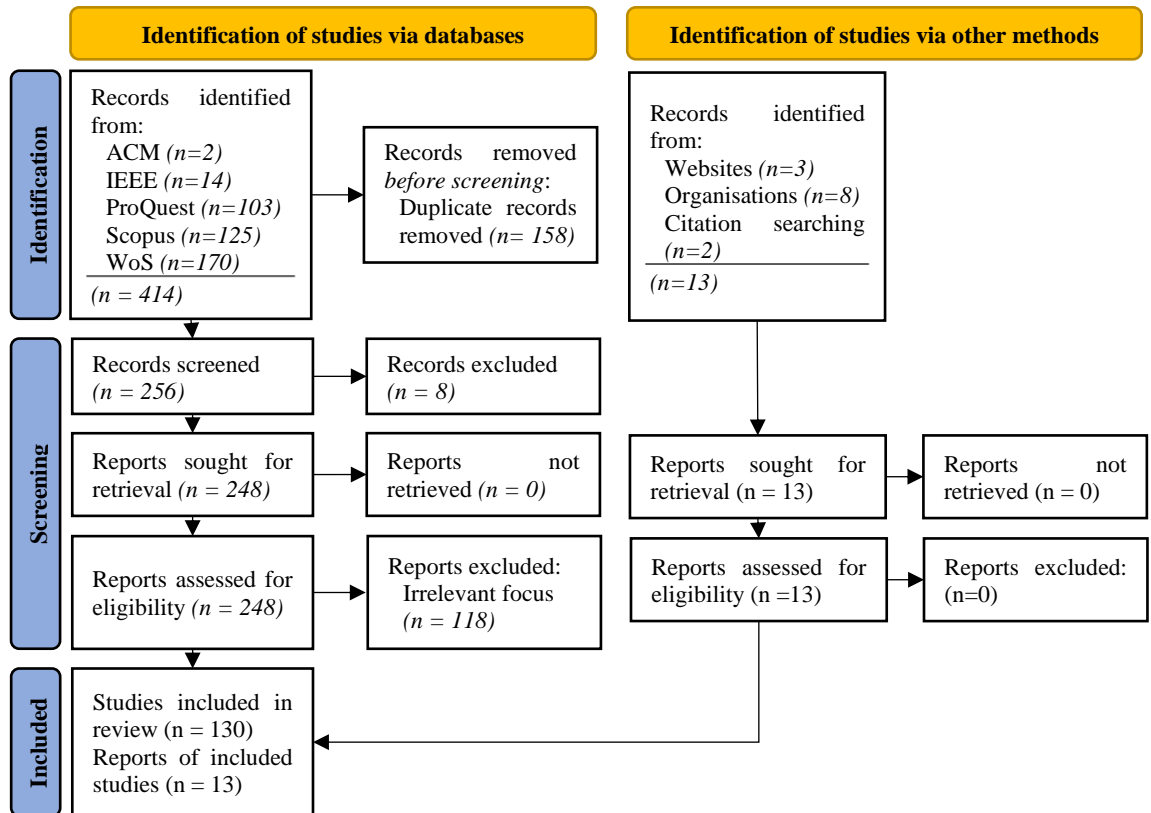

*Figure 6 PRISMA systematic literature filtering methodology [98]*

### 3.2 Systematic Literature Review evaluation

Search terms used in this SLR appear to have successfully captured the key features of the topic; *Digital Twin* is of course a crucial phrase for this search. Likewise covering *semantic/ontological/cognitive* components are essential to filtering results to ontologically focused papers. In a prior review, D'Amico intentionally excluded CDTws, owing to the lesser relevance to their review and the minimal number of papers at that time due to the term's infancy [6], however, given the prominence of ontologies in this review and the more developed relevance of CDTws, the phrase *cognitive* was considered applicable. However, *cognitive* was the biggest source of manual screening, as several articles related to neuroscience and brain cognition. Beyond the SLR, other notable reports and papers were used for defining DTws and the Gemini Principles and were thus excluded from the literature classification.

Additionally, the search terms sought to include all instances of a Gemini Principle identified. However, given the terminology of the Gemini Principles is not unanimously established in research, it was clear that other synonyms may have been used to describe the principles in some work. For example, terminology for *security* is well-defined and established, with *resilience* being the only identified synonym (albeit as a different but connected theme), meaning the review appeared to capture a broad security frame with these two terms. In contrast, terminology relating to *federation* (*interoperability, interconnectivity etc.*) is more loosely and diversely worded, resulting in a longer search chain relating to this aspect; simply searching "inter*" as a blanket term would of course capture all articles mentioning the internet. As such, terms were iteratively added to the search as discovered by the authors of this review, and there is recognition that some unique terms may have been missed by this net. This limitation does in itself highlight a possible future area to improve standard nomenclature for how the concepts underlying these principles are discussed.

Interrater review achieved 83% agreement (54/65) between reviewers on which articles to include or exclude within the review, which exceeds the 80% agreement threshold deemed acceptable by prior studies [100]. Sources of disagreement predominantly centred over whether ontologies were sufficiently mentioned (mostly revolving around the use of the phrase cognition), with a few disagreements over whether any of the Gemini Principles were addressed. In these cases, the authors discussed the reasoning and appropriateness of these acceptance and rejections and agreed decisions.

**3.3 Systematic Literature Review domain findings**

Figure 7 illustrates the quantity of articles within each domain. 67 articles were reviewed in *Manufacturing*, 39 in *Built Environment*, 18 in *Other applications* and 6 *General* articles. Figure 8 in the Appendix shows that ontologies have been applied to myriad contexts and addresses unique challenges. As such, grouping research into these 4 domains clearly dilutes the nuance from the unique applications considered by individual articles. However, this classification does capture the broad idea that most papers focusing on ontologies to support a Gemini Principle for DTws are used within *Manufacturing* or *Built Environment* spheres rather than domains such as healthcare or energy, which even clumped together into *Other Applications*, pale in size as a category. This indicates that *Manufacturing* and *Built Environment* are leading the way in recent advances, which should be unsurprising considering that DTw have their roots in these domains through the precursor technologies of PLM and BIM. Additionally, both domains also depend on broad cross-sectoral supply chains, potentially indicating more viable business cases. Moreover, only a few papers within this review were purely theoretical or *General* in their application, reaffirming that industrial applications may lead recent focus on applying ontologies. Prior literature suggests that the number of articles relating to DTws has greatly increased over recent years [16], and is certainly in line with the findings of our work; 6 articles were published in 2019, 13 in 2020, 32 in 2021, and 49 in 2022. Whilst only 17 were published in 2023, the year is ongoing at the time of this review. Interestingly 0 articles reviewed predate 2019, apparently confirming the relative infancy of the particular vein of research captured by these search terms.

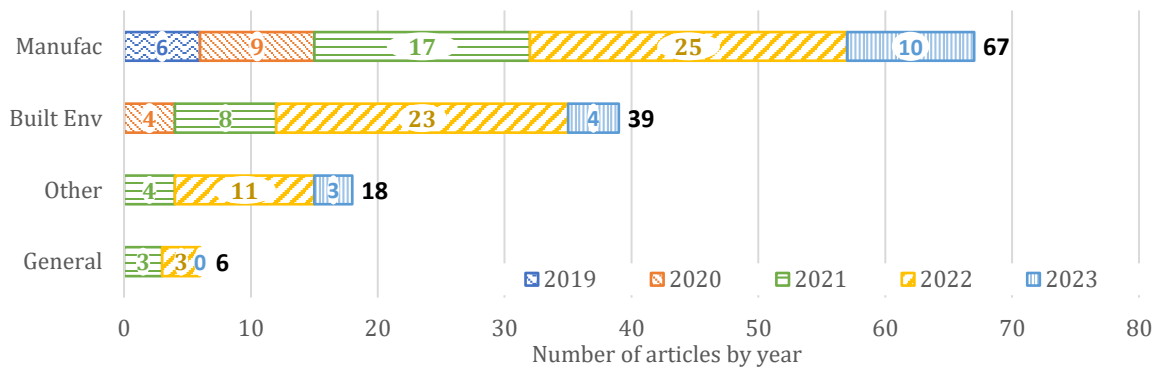

*Figure 7. Number of systematically reviewed articles from each domain*

Whilst the Gemini Principles are *Built Environment* focused, subsequent work notes that cross-domain analysis may be warranted to futureproof DTws beyond individual domains; the Apollo Protocol is an ongoing set of guidelines for cross-sectoral DTw collaboration between the built environment and manufacturing [101]. As such, research gaps for this study seem to split into 2 fronts: 1) furthering ontological capabilities within more mature *Manufacturing* and *Built Environment* sectors. 2) adapting existing ontologies and findings from *Manufacturing* and the *Built Environment*, into *Other*

*Applications*; individual surveys do exist for some of these sectors (many of which included within this review) and could be expanded with more practical ontologies to address the particular criteria and concerns of these less mature applications.

## 4 ONTOLOGIES FOR THE GEMINI PRINCIPLES

Whilst ontologies are not explicitly considered by the Gemini Principles, there do not seem to be any obstacles to incorporating in this framework. Indeed, a review of the literature seems to establish that the strengths of ontologies can support DTw observance of each of the Gemini Principles requirements. This section explores how current implementations of ontologies within literature is supportive of each of the Gemini Principles in turn.

### 4.1 Lens A | Purpose

The first lens of the Gemini Principles focuses on ensuring the purpose behind the DTw is clear, to ensure that use is beneficial to both users and broader society, contributing to the *Public Good*, producing *Value* and revealing *Insight*.

*4.1.1 Principle 1 | Public Good*

Ontologies can contribute to the *Public Good* by mapping out metrics for DTws that are important to a wider range of stakeholders or support properties such as safety and sustainability. This is important for ensuring that the implications of DTw models on the wider world can be transparently documented and assessed.

1. **Catalogue stakeholder engagement:** DTws currently seem to be viewed as primarily technical systems [51]. As a result, social or ethical implications (i.e. who is affected and how) ensuing from DTw design or implementation choices can be overlooked [66], [97]. Despite this, DTws could be capable of bring societal and economic benefits to people globally (e.g. tackling citizens problems in a smart city) [4], [10]. DTw owners can mitigate these issues by engaging a wider range of stakeholders directly or indirectly impacted by the asset. However, stakeholders often have many diverse and conflicting perspectives. Interpreting the priorities raised by a wider consultation, particularly for interconnected, shared assets (e.g. cities) has the potential to introduce more complexity into the DTw. Ontologies can map and resolve conflicting demands and competing objectives [75], to more feasibly engage with a broader spectrum of stakeholder perspectives.
2. **Improve safety modelling:** Whilst DTw deployment and simulation in isolation can support safe working and early warning systems [3], identify hazards [102] and maximise user wellbeing [103], this may not fully consider humans and machines interacting in the same space. In that regard, ontologies can improve DTw-based approaches to safety management and embed considerations of safety across design and operation [14] as well as semantically modelling hazards and defining safe use case parameters in human and machine readable terms [102], [104].
3. **Map sustainability metrics:** Current DTw applications to increase asset efficiency through reduced waste and greater reuse can already advance sustainability and the circular economy (wherein waste products of one process are inputs for another) [3], [55], [94], [105]. Furthermore, DTws can be applied to address goals set by local communities and global UN Sustainable Development Goals (SDGs) [94] through capabilities to: reduce energy use and increase energy grid flexibility (Goal 7: Affordable and Clean Energy), link assets to produce innovative insights and increase system resilience (Goal 9: Industry, Innovation and Infrastructure), enable participatory democracy by engaging citizens in planning and development (Goal 11: Sustainable cities and communities), and improve asset system efficiency, reduce carbon footprint and quantify costs of climate inaction (Goal 13: Climate action) [3], [81], [94]. Ontologies can further these efforts by processing the appropriate metrics for the DTw to embed sustainability or SDG guidance across the whole asset lifecycle, once stakeholders map such goals [43].

*4.1.2 Principle 2 | Value Creation*

Ontologies support *Value Creation* in DTws, through establishing new data links and metrics in a human and machine readable way, supporting optimisation and automation. This could allow DTws to further exploit existing data, guide data-driven decisions and make more effective use of resources.

1. **Deduce data links:** DTws may maximise value generated through greater real-time connection and modelling of assets across the whole supply chain and lifecycle to help stakeholders to share and use greater amounts of data [2], [3], [55], [66], [76]. Ontologies could more efficiently link existing data and deduce new useful connections, to maximise value obtained for businesses using DTws across design and operation stages [14], [94], [106]. Linking data for cross-asset purposes and applications could lead to more useful inferences from sensor data, significant asset performance improvements (improved cost, time and efficiency) and novel business opportunities [2], [9], [40], [44], [55], [64].
2. **Define multiple value metrics:** DTws are also an opportunity to better define sources of value, which may currently be disregarded (i.e. unidentified changes that can be made to reduce costs or increase sales). Whilst DTw adoption may in itself be a necessary edge for general competitiveness and productivity, the distinct sources of value may need defining [3], [94], as financial and human barriers and incentives to implementing DTws can vary by sector or use case [2], [3], [6], [19]. As such, there is no single business model for DTws [20], meaning different user relationships can exist (such as owning or renting DTws during necessary stages) [2]. The most cost-effective level of detail within DTw models also needs determining; vaster, more realistic and interactive DTws are feasible as computing components become more advanced and cheaper [44] and economy of scale can be gained via multiple DTws on the same server [64], however larger companies may be more prepared to maintain the necessary IT infrastructure [107] (though it is true that cloud servers have a quasi-unlimited capacity for hosting multiple DTs). As such, optimal solutions are circumstantial, requiring stakeholder coordination along supply chains [6] to tailor what data should be captured, analysed and shared to maximise value [46]. Any attribute may be measurable once connected to the DTw system [10]; breaking a top-level business problem into sub-questions down a hierarchy of sub-DTws could support decision-making by helping users visualise various quantitative and qualitative Key Performance Indicators (KPIs), trade-offs and constraints [88], and could evidence the effectiveness of the DTw in different use cases [11], [13], [49]. Ontologies could unlock new sources of value for DTws by defining and structuring value metrics and KPIs, then linking relevant data for complete performance analysis across the whole DTw system, and flexibly responding to changing business requirements as indicators are adjusted [43], [108].
3. **Support automation and optimisation:** DTws are already capable of optimising assets, by adjusting asset components and parameters to interact more efficiently with other components and the environment, [38], [43], [78] to reduce waste and increase productivity. However there are challenges in quantifying improvements, working with multiple assets concurrently, modelling more complex environments and computing large data sets [2], [3], [79]. This is particularly challenging for a DTw system of systems (i.e. an entire supply chain) as each asset can have components with different functions, data scales and KPIs, whilst vast data collection is needed [9], [44]. Meanwhile, DTws also support automation, control of all or part of an asset with human involvement reduced or eliminated, of decision-making, to make operational choices that increase productivity and profits without human involvement, including managing facilities, guiding or controlling vehicles and producing schedules [3], [4], [78], [109]. Automation by its very nature needs fully machine readable data [110], meaning machine readability of ontologies potentially aids automation of decision-making and asset operation [27], [72]. Ontologies can add value to DTws by enhancing capabilities for asset optimisation and automation, and ontologies within CDTws could train ML to infer and predict from existing datasets for real-time automated control and decision-making [13], [79].

*4.1.3 Principle 3 | Insight*

Ontologies can generate *Insight* in DTws by contextualising and linking DTw data. This allows for deeper understanding of information to be given to DTws for new functionality and informed decision-making.

1. **Contextualise data:** DTws are becoming used for monitoring (observing assets through collected data) [58], [60], [72], [78], [111] to inform users how assets behave across the lifecycle and in real-time [13], [14], [71], so that more informed decisions and faster information flows can be achieved [48], [88]. However, raw data may only be meaningful within the context it is collected, particularly where sensors capture real asset behaviour. Therefore they may require processing and combining by the DTw to correctly reflect what is happening in real-time to bring decision-making insights of business value [3], [37], [73], [88]. Ontologies can bring context and meaning to data by mapping out relationships between stored information, which can then be graphically represented through DTws [65], [67]. DTw literature observes that ontologies can be used for simulation as: 1) *schema* (theoretical, example ontologies), 2) *real-world applications* (ontologies used with a physical asset, often as a database), 3) *aggregating external data sources* (ontologies creating simulation models by integrating different datasets), and 4) *scenarios* (simulating a scenario wholly through ontologies) [112]. In particular, ontologies within CDTws can deduce insights through a reasoning engine (a component for analysing metadata to deduce new insights and patterns) [113].

2. **Link data across asset lifecycle:** DTws may be created over a long timeframe, playing various roles for different data and changing information needs, challenges, priorities and intentions over several lifecycle stages [13], [31], [51], [65], [106], [111] and are already capable of bringing out high-level information (which provides a broad overview) about assets across planning and operation stages [20], [58], [72]. As such, DTws may become more valuable as they are deployed over longer stretches of the asset lifecycle [6] and DTws for different lifecycle stages could be integrated to be fully comparable, using different overviews and data linkages at different stages [2], [13], [31]. This could lead to a more profound understanding of assets by enabling communication across supply chains that interact at different lifecycle stages [33], and supporting predictive maintenance (using past data of failures to predict when a functioning component or asset is close to failure) to reduce downtime and cut down operations and maintenance costs [6]. Ontologies can create insight by temporally (relating to time) connecting data across the asset lifecycle. In-depth ontologies can define component interactions for particular stages [33], [43], while a general ontology can support a DTw for the full asset lifecycle [33] which could learn from past data to infer the present context and improve estimations, decisions and predictions [2], [3], [79].

## 4.2 Lens B | Trust

The second lens of the Gemini Principles conveys the aspiration for users and society to trust DTws, both in terms of outputs, and protecting data use, through improved *Security*, greater *Openness*, and higher *Quality*.

*4.2.1 Principle 4 | Security*

Ontologies can improve DTw *Security* by identifying and responding to threats, denoting access control and facilitating Distributed Ledger Technologies (DLTs). Ontologies are important for ensuring consistent security management of DTws, allowing secure best practice to be more universally embedded.

1. **Threat identification:** DTw platforms (software and hardware), assets and generalised data may need securing to prevent unauthorised access, mitigating cyber threats that leave DTws vulnerable to industrial espionage, stolen data, and sabotaged equipment. This risk is worsened by new potential threats and attack surfaces (weak points in system security) arising as DTws become more established [3], [19], [44], [64]. To address vulnerabilities (security weaknesses) if communications are intercepted (data seen or altered by the wrong person), devices connected to assets may need identification, authentication (to verify identity and provenance), trustworthiness evaluation and, if these precautionary measures are satisfied, authorisation and granting of controlled permissions. Ontologies can support security policies (guidelines for protecting systems and information) that evaluate and prevent security breaches in DTws, then can evaluate potential attack impacts [87].

2. **Incident response:** DTws can track device and system security, and monitor for unusual behaviour, verify when users perform a particular action, model attack scenarios then enact appropriate countermeasures against likely future attacker behaviours, and inform users of faulty or compromised devices [44], [73], [114], [115]. Ontologies can then facilitate the assessment of users or devices against security attributes prior to allowing access to the DTw [114]. They can subsequently also monitor device state, location and time after a device is used [114].

3. **Denote access control:** Assets segmented into subsections for security could be mirrored by a DTw with sub-DTws [110] of different access criteria. However, partitioning assets could lead to data silos (internal to the DTw or external between stakeholders). Ineffective data transfer between silos could present security risks and limit knowledge gained by combining data [1], meaning there is a balance between ensuring that DTws and the data stored are fully secure yet accessible in real-time [66], [74]. Additionally, stakeholders may generally refrain from sharing data or access, due to the lack of capabilities to selectively share between organisations, alongside privacy, intellectual property, ethical and regulatory implications of sharing data that is either confidential or can reflect user behaviours [9], [51], [64], [76], [94]. Unauthorised access could lead to data deletion, use or change, so it may be desirable to enact measures to ensure important data is appropriately immutable (unchanging with time) and encrypted (obscured so only users with a passcode can see), however this is difficult when sharing across stakeholders with different IT infrastructure [1]. Role or attribute-driven controls could grant selective access to full data or specific attributes [64], but it may be unclear which data is openly accessible or confidential unless this is actively considered when data is structured [51], creating an argument for security descriptors in DTw ontologies. Ontologies can then enforce security policies to improve security, privacy and data integrity [73], and are capable of evolving (adapting with time and innovations) in conjunction with changing security requirements of an ever-shifting threat landscape [37].

4. **Support Distributed Ledger Technologies:** For DTws to be able to manage the security of assets and data, it is important the DTw itself is secure [55]; web-accessible DTws must incorporate any usual security requirements for online devices. Yet further security features may also be needed for critical DTws [64]. Sharing data from DTw systems via Distributed Ledger Technologies (DLTs) or blockchain (distributed protocols where data is spread across many locations securely) that can handle vast amounts of data, can: 1) improve security, transparency and data integrity by signing (checking a user genuinely requested information) and encrypting (so only the authorised entity can decipher and read it) data; 2) increase privacy by reducing the amount of data transmitted, and supporting options for data sharing which is public (open to anonymous clients), or private (with permissions and predefined entities); 3) expand scalability (ability to work with increasing demands as system grows) by providing immutable (unalterable), structured, distributed and secure storage and capabilities to handle large amounts of data with access and sharing rights (to control who can use and share data) [1], [31], [62], [77], [116]. However, DLTs may be unable to identify or authenticate users due to the large quantity, complexity and heterogeneity of devices [117]. Ontologies may synergise with DLTs by structuring data to adhere to data governance policies to support data reuse and protecting access to sensitive data by controlling how data on DLTs can be retrieved and by whom [1].

*4.2.2 Principle 5 | Openness*

Ontologies support the *Openness* of DTws by linking data across silos and mapping out FAIR Principles, in a human and machine readable way that can synergise with internet infrastructure. This is important as ontologies are most useful when shared and reused by an ecosystem of different parties, to maximise collaboration and data accessibility across sectors.

1. **Enable Open DTws:** 'Open' DTws 1) allow instant, free, modifiable implementation: being able to edit the DTw source code, redistribute DTws across manufacturers or sectors and seek support from online communities, can all speed up prototyping. 2) remove vendor lock-in: eliminating software license costs and extending DTw lifetime by removing dependencies on a company software for updates, maintenance or third party extensions. 3) follow open standards: facilitating interoperability, even amongst closed DTws. The result might be a quicker design cycle, interoperability between lifecycle stages and a less fragmented supply chain [9], [44], [64], [67], [81], [118]. Whilst stakeholders may support reducing silos [81], many DTws are currently closed (unconnected to the internet), proprietary (company-specific) systems, with architectures of differing definitions and relationships that can lack the above benefits of open, shared systems. Data silos in DTws can also arise from software used, devices adopted, and datasets from isolated domains, as well as a lack of programming knowledge or motivation to connect DTws to the internet [1], [2], [6], [9], [44], [64], [65], [106]. Nevertheless, an open DTw can incorporate other closed DTws [10]. Ontologies could overcome DTw silos by linking data from different systems. Ontologies are inherently shareable (being modular and using a standard data format) [37] and can support data integration and reuse in an open, visually clear way, for heterogenous devices across silos [43], [114].

2. **Embody FAIR data principles:** Data fragmentation can prevent stakeholders from benefiting from sector specific knowledge, access to real-time information, or insights from connected data sources [27], [81]. Ontologies can overcome data fragmentation and embed FAIR (Findable, Accessible, Interoperable, and Reusable) principles into metadata. Adding publicly visible metadata descriptions to DTw data may make data more searchable and support model reuse in similar contexts, reducing development time or the need to repeat tests to reobtain results; however current metadata approaches may not be scalable or user friendly, which ontologies could address [64], [73], [107], [119]. 'FAIRness' can be stored as metadata [32], allowing ontologies to use real or simulated metadata descriptions of device capability and properties to link data, relationships and interactions [43], [110], [114].

3. **Facilitate HMI:** Structuring DTws to be readable and interoperable between both human users and devices, DTw models, interfaces or information exchanges, would improve system efficiency [8], [10], [20], [29], [73]. Ontologies can provide this common structure understood by both humans and machines, which makes information machine readable, and extractable by sharing and reusing knowledge from heterogenous data [27], [108]. However, terms and descriptions used within the ontology may need sufficient richness (using more descriptive and detailed language) to create semantic relationships that are capable of dealing with complexity whilst supporting real-time decisions and remaining both human and machine interpretable [74], [89]. This could potentially use natural language definitions (words and sentences as spoken by humans, and used by e.g. chatbots) to prevent misinterpretation or the use of ML training [46], [65].
4. **Facilitate internet infrastructure:** Ontologies can synergise with web services and Application Programming Interfaces (APIs) (interfaces for software to interact) to support open DTws. APIs often act as middleware communicating between applications, linking open source software [9], [47] and could oversee communications between devices and mitigate incompatibilities between different communication protocols (rules defining data exchange between devices) [109]. APIs can communicate with and alter ontologies, as well as using the ontology to process data external to the asset [2], [112], with the potential to support quicker, more stable, adaptable and more realistic DTw models by storing system configurations in a database [24], [120].

*4.2.3 Principle 6 | Quality*

DTw *Quality* can be improved by ontologies which can evaluate data quality, and adapt data fidelity to what is needed. Using ontologies for high quality data matters to ensure a correct and accurate DTw.
1. **Work with big data at high precision**: DTw models can be high-accuracy, and can pair with precise sensor data to quickly and reliably reflect and minutely adjust asset parameters and outputs (e.g. quality of a manufacturing product, or modification of a component for maintenance) [3], [36], [41], [42], [72]. Yet the accuracy of DTws rests on the quality of new data, visualisation quality and the appropriateness of assumptions made in the model [3], [55]. This begs the question about quality metrics for DTws. As such, data quality needs measurement to ensure users make decisions with DTws that are grounded in reality [74]. Furthermore, less precise or incorrect data or results can be difficult to quickly identify in large data sets, and can result from discrepancies in: how structured a dataset is, the scale of data collected, the quality of algorithms used for analysis and decision-making, the extent of historic data collected across asset lifecycles, and whether data originates from third party sources without clear provenance (record of data origins and history) [2], [9], [15], [41], [107]. As a result, most data may not currently be reused [43], despite potential cost and time savings in reusing existing models (i.e. not having to recreate models or re-collect data) [15], [73]. The sheer quantity of data collected and sources of value within DTws may justify expending effort to fix this [27]; yet manually labelling mass quantities of images or videos is clearly impractical, although it can mitigated with a semantic approach to categorise images and videos for humans or machines [80].
2. **Map quality metrics:** The range of data, scenarios and devices [83] means no set singular judgement for quality, usability or suitability of DTws exists [118] including: whether *quality of user experience* (response time, rendering issues) or *quality of service provided* (delays, lags) are more important [77], or deviations between individually measured *As Designed* components (which may not physically exist) against measurement of *As Built* components (within a finished asset) [3], [39], [121]. Pairing ontologies with services could identify data quality through rule-based analysis to: monitor and learn appropriate behaviours, detect anomalies for improved future fault prediction, provide a communication interface with up-to-date information between DTws, optimise DTws, and make decisions [69]. Ontologies can also work with incomplete, inconsistent or high uncertainty knowledge, and iteratively refine relationships and patterns as new data is discovered [75], [118]. Increasing the data richness by adding metadata and linking historic data [9] could further support evaluation of model completeness and ontology consistency [119]. Additionally, ontologies can support measurement of data quality in terms of correctness (how many values are erroneous?), completeness (how many values are missing?) and accuracy (how close are values to what they should be?) [121].

3. **Refine data fidelity:** Ontologies can support fidelity management (accuracy and precision of information appropriate to a context), allowing data simplification. DTws face challenges in integrating data of differing scales (e.g. more detailed localised data with broader data categories with wide reaching data) for different processes, including data and models used outside of their intended collection purpose [59]. As DTws are exposed to larger datasets from real-time data streams, exacerbated by cheaper sensor networks, faster project lifecycles and greater data variety [34], [110], [122], integrating data and systems becomes exponentially complex, growing increasingly fast as size increases [65]. Yet, data criticality can vary (i.e. not all data is needed, or may not be needed to the precision collected) [36]. As such, computational resources, asset efficiency and project value could all be optimised [43], [72], [119] by modelling key data: 1) as simply as possible and proportional to scenario complexity (i.e. a quick estimate, or a more in-depth analysis) [9], [73], 2) to meet a certain accuracy level or response time required by a particular model, device or scenario (e.g. purpose may need real-time or just periodic updates for Quality Control) [9], [82] 3) balancing accuracy against technical and financial costs of collecting and using data [36], [39], [123], 4) adding additional parameters only as the environment is validated [119] (i.e. so that the number of unverified sources of uncertainty is controlled). Complex DTws can be broken down into different granularity scales [43], [93] with switchable fidelity in models that can be simplified and aggregated [123], then reused in particular scenarios and structured using an ontology [36]. Synergising more permanent static data (that does not change) with more reactive dynamic data (that changes in real-time) may be a particularly salient use case. However, different datatypes have different information needs and change between lifecycle phases (i.e. static data should be updated less frequently to reduce storage and computing requirements) [36], [45], [106]. Ontologies can standardise processing for heterogenous data of multiple datatypes, scales, and granularities (levels of detail), then support data cleaning (removing missing or incorrect data) and packaging (formatting and storing data) by mapping correctness of syntax (grammatical structure) and provenance (traceability of prior data uses and origin) [17], [37]. Ontologies can also define high and low fidelity models as well as the relationships between these, for more flexible precision in the level of detail shared with stakeholders, supporting more advanced functions and managing Big Data at workable levels of detail [32], [110], [123].

**4.3 Lens C | Function**

The third lens of the Gemini Principles has a focus on ensuring DTws can Function alongside other DTws and innovations, now and in the future, via *Federation* of the DTw environment, suitable *Curation* practices, and capabilities for *Evolution*.

*4.3.1 Principle 7 | Federation*

Ontologies can support *Federation* of DTws by mapping relationships between heterogenous systems, particularly when structured around standards or reference architectures. This is important for different DTws and ontologies to seamlessly interoperate, understand each other, and be combined or compared.

1. **Translate heterogenous systems:** Managing DTws individually may be appropriate for isolated assets or subsystems, but shifting to managing a DTw network using a systems approach, can support big picture decision-making [88] by assimilating data of different formats [73]. However DTws that may interconnect within problem domains could significantly differ in scope or scale to DTws from other sectors [2]. Yet, whilst many DTws may not currently be fully interoperable [51], [55], integrating systems can support collaborative data-led asset management and transparency across lifecycle stages and between sectors [22], [73], [80]. Linking DTws into Federated networks of DTws, can result in inter-DTw learning, interoperability and integration [6], [35], [55]. A single DTw system could include multiple DTws for different components, asset subsystems and lifecycle stages, each built by different sectors with different standards, architectures, and semantics [12], [124]. Interoperability issues from heterogenous data [35] of multiple sources [34] may increase as devices become more complicated and diverse [72] potentially resulting in inconsistently modelled DTws, hindering interoperability, reusability and information sharing [33]. Ontologies can support interoperability for compatibility between systems from different vendors [20]. Additionally, ontologies can unify information between heterogeneous DTws across sectors without irreversibly transforming data, so new DTws can be easily added and integrated to a hierarchical DTw system at the appropriate level (i.e. a component as part of an asset), whilst preserving asset autonomy and constraints [35], [48], [80], [108].

2. **Interpret reference architectures:** Common ecosystems (connected data environments) and their associated data governance policies (rules around data use and collection) may need to integrate knowledge across sectors or fragmented, complex environments [57], [125]. Reference architectures are standardised frameworks for common implementation [52] of the requirements, functionality, relationships, principles, structure and mechanisms needed by a DTw [124]. Ontologies can link different reference architectures through common frameworks, data format and structure to allow communication between components and systems [124], resulting in collaborative ecosystems of DTws that follow asset processes rather than isolated tasks (e.g. a manufacturing line) [88]. However, a single ontology may not always be possible or desirable given that the context a DTw is implemented in can contain other sub-contexts, meaning different DTws may have different needs and could thus require contextual ontologies [35].

3. **Enforce standards:** Beyond reference architectures, ontologies can follow standards for various components and processes. The lack of universally agreed standards across all DTw use cases or lifecycles means data may be differently modelled or structured [52] limiting integration of these DTws [10]. However, components connected to most commercial DTws may already follow existing industry standards, supporting interoperable data exchanges between these [20]. Thus it is the relations between DTws that still needs to be standardised [64]; differing DTw standards may need blending to cover different sector specific challenges [90]. Furthermore, linking and standardising DTws mean that existing DTws could be benchmarked (using performance of example use cases as a comparison) [11], [14]. Ontologies can connect data, assets and models [8], [66] across applications with different standards and requirements [7]. A universal standard could form the foundation for a common top-level ontology, to ease integration of DTws from across sectors as the same overall dictionary is used [6]. Structuring ontologies around open standards allows reusability between different software and stakeholders [67], [93]. Ontologically linked DTws can integrate diverging standards from multiple bodies to combine each standards' expectations; some semantic interoperability may already exist between common industrial standards *IEEE 1451* and *IEC 61499* [7], [90].

*4.3.2 Principle 8 | Curation*

Ontologies support *Curation* of DTw data through being extensible, enforcing data governance policies, and linking historic data:

1. **Improve extensibility:** Clearly defined and agreed assets and data reduce ambiguity in DTws' documentation [3], [88]. Ontologies can provide defined groundwork for DTws and can attribute a common correct version of knowledge that is extensible (can work with new functionality) and scalable (can work with larger size systems), to bring shared unambiguous meaning [35], [106] understandable to both humans and machines [32].

2. **Administer data governance policies:** Literature suggests that DTws may be needed to digitally formalise processes (explicitly structure rules and requirements) [48] and adopt data and asset governance (management of data policies and assets) [60], [61] to record and resolve different prioritisation of factors by different users at different stages of the asset lifecycle [93]. By mapping out a transparent, open set of governance rules, ontologies can integrate resources to create shared, interoperable knowledge with improved traceability and decision consistency [88]. In particular, CDTws could be deployed and operated across the supply chain, with stakeholders jointly configuring the CDTw at the design phase: the information to include; the level of decision-making autonomy to be granted (including identification of business, operational and ethical decision-making considerations), and the allocation of liability and accountability for data processed from different owners along the asset supply chain [13].
3. **Standardise terminology:** Historic data about assets and DTws may need efficient and traceable storage to be useful for real-time analysis that is fully comparable across the entire lifecycle for the different overviews and linkages needed at different stages of the lifecycle [11], [37], [41]. However, definitions used for assets or components can vary within an organisation [11]. As such, using ontologies to define a common framework and terminology could make data more reusable, and facilitate data sharing [8], [94]. Furthermore by exploring relationships between data and providing context, ontologies improve the interoperability and reusability of data [65], [67].

*4.3.3 Principle 9 | Evolution*

*Evolution* (capability to change with time and new technologies) of DTws is supported by ontologies which are able to manage version control and are suitably structured to be amendable for future innovations and to incorporate AI. This ensures ontologies remain a relevant tool and are adaptable to future changes.
1. **Versioning: DTws** may need to be able to evolve and respond to potentially time-sensitive real-time updates in ever changing assets, however, overwriting existing data can cause regression and prevent its reuse, unless different versions are saved [37], [41]. Keeping different versions for data provenance (record of data origins and history) could be particularly problematic in a scenario where various stakeholders in a project may create new versions that may conflict with prior versions [126]. Ontologies can manage the collation and merging of information changes and model versions, storing parameter states (current values and conditions of different asset properties), with new versions created as these change, and can provide metadata contextualisation for autonomous interoperability and version information [37], [43], [73].
2. **Adapt with improvements:** Design, configuration, control and operation of DTws or assets will continually evolve with new technologies, methods, tools, processes, materials, use cases, operational and business strategies and become increasingly complex with more connected components and subsystems [12], [42], [82]. Ontologies could support adoption of different data sources [57]: 1) rich semantic information for a particular project (e.g. from BIM) or, low detail, wider contextual information (e.g. from Geographic Information System (GIS)) [108], 2) semantically represent knowledge to support Augmented Reality (AR) and Virtual Reality (VR) [2] visualisation of DTw models [16], or 3) communicate directly with Unmanned Aerial Vehicles (UAVs) [87] or IoT sensors [90].
3. **Incorporate Advanced AI:** Ontologies in CDTws could support self-learning systems (trained with historic data to improve performance over time) [48], using Artificial Intelligence (AI) to coordinate and adapt to changes and model potential future scenarios [22], [82]. However literature suggests the term is used imprecisely [4] and comprises algorithms with purposes of data mining (extracts patterns or insights from datasets), ML or Deep Learning (DL) (a form of ML using layered neural networks to simulate human brain function for more advanced tasks [127]) , that can each be used with DTws [107] to further link data and maximise business advantages [11], [95]. Literature suggests ontologies can pair well with AI, and can bring clarity to decisions made by ML-trained systems [14], [75], [128], [129].

4. **Embed modularity:** As many devices are used across various lifecycle stages, a single simulation is unlikely to be able to process and convey all information in a timely way. This suggests that modular (separated into individual sections that can be individually added or changed) DTws may be beneficial [23]. Indeed, rather than a single system, DTws consist of sets of infrastructures [10], meaning, modular architecture made of microservices (services with granular functionality but not necessarily minute code) can simplify DTw code, make it easier to maintain, support decentralised architecture and can be created by different teams using various coding languages, improving flexibility and scalability [9], [25], [73]. The modularity of contextualised ontologies mean these are easily maintainable and reusable in existing or future contexts [35], [73].

## 4.4 Summarising use cases – how useful are ontologies in meeting the Gemini Principles?

Ontologies have been shown to be relevant for each Gemini Principle.

A) **Purpose: 1)** Literature found ontologies can contribute to the *Public Good* through engaging a broader range of stakeholders, improving safety monitoring, increasing stakeholder engagement, and governing sustainability metrics. Possible further areas not mentioned by literature could include ontologies for identifying and overcoming data biases. **2)** Literature indicates that ontologies could create *Value* by creating profitable connections, providing metrics of value, as well as supporting automation and optimisation of assets. **3)** According to literature, ontologies could generate *Insight* through contextualising data, and linking data across the asset lifecycle.

B) **Trust: 4)** Literature highlights that ontologies can improve *Security* by identifying threats, responding to incidents, supporting dynamic access controls, and synergising with DLTs for secure data storage and sharing. However, literature did not discuss how well ontologies mix with existing security measures, nor how to ensure that the ontology itself is secure. It is important that ontologies should not introduce a vulnerability, as any DTw system needs to be secure throughout. **5)** Literature studies found that ontologies can support *Openness* by supporting Open DTws, facilitating FAIR data management, facilitating Human Machine Interaction and integrating services, APIs and web technologies into DTws. **6)** Findings from literature note that ontologies can improve *Quality* by mapping quality metrics, and dynamically managing data fidelity. Yet, there was no mention of evaluating the quality of the semantics of an ontology itself, nor how ontologies could impact the quality of service provided by DTw.

C) **Function: 7)** Literature highlights that ontologies can support *Federation* by managing interoperability between heterogenous systems and enforcing common reference architectures and standards. **8)** Literature found *Curation* ontologies could support extensibility, manage data governance policies, and reuse historic data. **9)** Literature reports ontologies can sustain *Evolution* by tracking version changes, supporting modularity, and integrating new innovations and AI. Yet, interactions within ontologies between new and legacy innovations were not discussed.

Regardless of the intended outcome or lens, many of these opportunities appear to amount to using ontologies to link devices and data to enable interoperability, or for governance policies to decide how data or devices are handled. Whilst it could be argued that applying a Gemini Principles overview is therefore superfluous, instead, the use of the Gemini

Principles demonstrates the wide range of applications that embedding ontologies into DTws can bring, whilst allowing evaluation of ontology implementation to be enmeshed into evaluation against wider DTw goals.

## 5 CONCLUSION

Literature reviewed in this SLR encompassed research which used ontologies to non-explicitly address each Gemini Principle in turn, primarily to federate networks of connected DTws and to map important relationships and properties as metadata, pointing towards research directions in furthering these topics. Most literature originated in Manufacturing and the Built Environment; these are more developed sectors with more mature DTw ontologies that can continue practical work to build deeper ontological interoperability. Meanwhile, other less developed sectors such as healthcare or energy could benefit from adapting successful work from these more mature sectors, to align with sectoral requirements and context.

### 5.1 What are key areas for future work?

Existing literature indicates that specific future directions of DTws are hard to predict, but could be highly impactful [2]. As DTws are interdisciplinary, there are areas for advancement through research, alongside challenges for industry and government to resolve, gathered in the subsections below (though many points raised could require input from all 3 groups).

*5.1.1 Digital Twin gaps in research*

Literature has identified several research gaps in representing unconventional properties and considering the future of DTws that require further research:

1. **Representing unconventional properties**: Gaps may exist in ontologically representing real world ambiguity [15], and more abstract concepts (including actions such as pushing or pouring) [72]. Once ontologies are more sophisticated, applications of DTws for economic modelling could also be investigated [47]. One possible solution is to use Natural Language Processing to guess links between assets and data for an ontology [9], [31].
2. **Futures of DTws**: Research on how ontologies will evolve is in early stages [37]. Yet, challenges in generalising existing enabling technologies [34] mean integrating DTws is difficult due to: system complexity, unfeasibility of aggregating large datasets, lack of historic data, continual changes in some operating parameters and conditions, and complexity in handling data ownership [1], [2], [23], [31]. Furthermore, capabilities to easily create a top-level DTw for an individual asset may be needed before attempting to make networks of DTws [64]. Whilst ontologies are extensible, there is a research gap in ensuring ontologies are structured in as reusable a way as possible.

*5.1.2 Digital Twin gaps in industry*

There are technical challenges industry may need to consider including consideration of the full DTw lifecycle, wider security and privacy of the DTw system, interoperability across sectors, and overcoming barriers to adoption:

1. **Full lifecycle consideration**: A technical challenge could be managing an ontology for the entire lifecycle of a digital twin system, (from design to maintenance). Standardised ontological languages from earlier stages of DTw lifecycles are often not used further or connected to other models [14], as the way ontologies are integrated depends on the particular task [80]. Furthermore, there may be a gap for expanding semantic attributes to work beyond the DTw lifecycle for other complementary tools such as BIM or GIS, with a possible need for formalised syntax [52].

2. **Wider security and privacy**: Security, privacy and safety are areas for future work [73] that may need to consider scalability, data loss, and human interaction [79]. As DTws integrate new approaches and technologies, new security risks or vulnerabilities may be created [9], [10] which may need defining and mapping by ontologies. Additionally, ontologically-linked KPIs may need to include metrics that consider security and privacy [9].
3. **Cross-sector interoperability**: Sectoral challenges in interoperability, flexibility and modularity [9] may remain, despite advances in ontologies. Further challenges for DTws lie in making DTw architecture scalable, adaptable, reusable and flexible [15] and to account for manufacturers using different devices and APIs [76], which could be resolvable through ontologies. Additionally, standardised manuals with sector expert feedback could be combined to guide ontology construction [109], with specific use cases to benchmark (test performance) an ontology alongside generic modelling documentation [52]. There is thus a gap in developing standardised ontologies that can be applied across various sectors, whilst accounting for different functionality.
4. **Overcome barriers to adoption**: Gaps exist in overcoming technical, financial and perception barriers for data sharing [55]. Thus DTws may require culture changes from industry and stakeholders [78]. The complexity of DTws increases as data elements are added and more technologies are adopted [2], [4], [31], creating challenges for DTw asset owners in: storing vast amounts of data created, guaranteeing real-time connection to moving or remote assets, supplying the high computation power needed to maintain sensors, and training personnel to operate DTws [37], [78]. Financial challenges arise from costs of collecting the vast dataset needed [46], maintaining servers and big data management [6], [46], [64]. Furthermore, there is technical complexity to overcome in: processing heterogenous data, linking time-based data with context, controlling access to information, and creating services that can semantically exchange information and access multiple real-time data streams simultaneously [25].

*5.1.3 Digital Twin gaps in policy*

There are policy gaps relating to improving standards around DTws, regulation and consideration of IP and liability.
1. **Standards**: There are challenges integrating services [31] and in linking different DTw models by CDTw architecture [31], as alluded to in Section 4. The particulars of services and open standards may need deciding upon, in terms of how centralised or decentralised, and how government-run or market-run [10]. This is particularly challenging for Standards Developing Organisations to standardise across sectors [31]. There is a standards gaps for semantics between DTws and related innovations, such as BIM [51].
2. **Regulation**: Stronger action than simply recommendations through standards, may be needed. Regulation could be required to both define liabilities and steer towards positive outcomes to ensure DTw remains a public good [97]. This could potentially necessitate future work in applying standards to ontologies themselves [25]; multiple popular top-level ontologies exist, possibly necessitating standards between them [31]. As such it may be important to ensure that ontologies support (and do not breach) such regulations.
3. **IP and liability**: Ownership of digital assets has been explored by prior literature [47], yet this has not been mentioned elsewhere in the literature. No works reviewed refer to using to ontologies define intellectual property, ownership or use rights related to data, models, or the insights derived from them these as metadata properties. Likewise, none of the literature mentions data liability nor using ontologies to assign this. These are both important considerations, which may be addressable through ontologies.

## 5.2 Closing remarks

This literature review has captured the capability of ontologies to support each of the 9 Gemini Principles, predominantly through federating DTws and data, alongside data governance policies. Most research relating to ontologies for meeting each of these principles is in the fields of Manufacturing and the Built Environment.

Future work exists in further applying ontologies to the Gemini Principles, as well as transferring lessons learnt from more mature sectors to less developed sectors. Additionally, research gaps exist in using ontologies to represent more complex or abstract properties and to ensure DTws are futureproofed. Meanwhile, areas for industry to progress are creating interoperable ontologies across sectors and lifecycle stages, ensuring ontologies and related innovations are secure, and overcoming technical, financial, or cultural barriers. In addition, policy gaps need to be filled in developing standards or regulation as appropriate and considering IP and liability. By bridging gaps between research, industry, and academia, this article intended to set the stage for continued progress in leveraging ontologies for resilient, federated DTws.

## ACKNOWLEDGEMENTS

This work is supported by the UKRI Engineering and Physical Sciences Research Council (EPSRC) grants EP/R513143/1 and EP/T517793/1.

# 6    APPENDIX | LITERATURE CLASSIFICATION.

*Figure 8 Table of application domains of systematically reviewed literature*

| Context | Objective/Challenge resolved | Ref. | Year |
|---|---|---|---|
| ***MANUFACTURING*** | ***Applications relating to I4.0, production, and manufacturing*** | ***Σ=67*** | ***2019-22*** |
| surface mining | Develop DTw architecture to resolve weaknesses in surface mining | [3] | 2023 |
| Human-centric industry | Systematic literature review of how DTws and AR can support human centric transformation of industry | [39] | 2023 |
| Human-Machine Interaction | Impacts of humans in the loop on cyber-physical systems in smart manufacturing | [74] | 2023 |
| Human-Robot Interaction | Facilitate HRI through DTws and Deep reinforcement learning, with practical prototype for validation | [104] | 2023 |
| Metaverse | Proposes method to combine DTws with industrial XR and ontology descriptions | [130] | 2023 |
| Industry 5.0 | Develop safety management DTw with semantic reasoning approach for I5.0 | [102] | 2023 |
| Tool maintenance | Proposes CDTw architecture for maintenance and validates through an example | [84] | 2023 |
| Quality assurance | Proposes ontology to align software for zero defects in manufacturing | [65] | 2023 |
| Digital factories | Framework for contextualised ontologies with case study | [35] | 2023 |
| Agile supply chains | Governance framework proposed for CDTws | [93] | 2023 |
| Factory automation | Explores architecture and ontology-driven guidelines for modular DTws | [20] | 2022 |
| Manufacturing | Literature review summarising data-driven product design | [131] | 2022 |
| Augmented Reality | Uses a DTw to support XR for more efficient and accurate maintenance | [122] | 2022 |

| Context | Objective/Challenge resolved | Ref. | Year |
|---|---|---|---|
| Cognitive manufacturing | Develops an interoperability framework for cognitive digital twin in manufacturing | [34] | 2022 |
| Smart manufacturing | Proposes a framework for a DTw and ontology for full lifecycle management | [33] | 2022 |
| Industrial infrastructure | Proposes a framework for high fidelity DTws across heterogenous devices | [15] | 2022 |
| Industry 4.0 | Creates a library of ontology predicates for human-machine interaction, validated through a case study of an assembly line | [115] | 2022 |
| Supply chains | systematic literature review on how DTws support supply chain resilience | [57] | 2022 |
| PLM | Ontology architecture for data governance in Product Lifecycle Management | [43] | 2022 |
| Standards | Develops a reference architecture for interoperable standards | [124] | 2022 |
| Industry 4.0 | Proof-of-concept for customisable DTw for product analysis | [46] | 2022 |
| Mechanical products | Using ontologies to fuse data across sources for full lifecycle DT | [27] | 2022 |
| Factory floor | Ontology to align DTws with standards | [7] | 2022 |
| Cognitive systems | CDTws for intelligent sustainable manufacturing systems, with case studies | [82] | 2022 |
| PLMt | Systematic literature review of DTws to support change management | [11] | 2022 |
| industrial systems | Case study and simulated experiments of CDTws for industrial systems | [12] | 2022 |
| machining | Deploy a Cloud-based DTw for monitoring process quality | [49] | 2022 |
| Production | Devises a conceptual ontology-based simulation for production DTws | [112] | 2022 |
| manufacturing | Systems thinking approach to CDTws to design an ontology for decision-making | [48] | 2022 |
| Product development | Literature review examining the roles of DTw as well as complementary tools | [58] | 2022 |
| Product management | Develops agile DTw model that is customisable to support product experiments | [68] | 2022 |
| Industrial IoT | literature review of how digital shadows can be combined with a roadmap | [59] | 2022 |
| industrial processes | Develops a framework for a modular feature modelling DTw with a case study | [23] | 2022 |
| Industry 4.0 | Literature review and conceptual framework on manufacturing scheduling | [132] | 2022 |
| Equipment testing | Framework for Digital Twin testing for Large Complex Equipment Components | [41] | 2022 |
| supply chains | Example methodology of CDTws for agility and resilience of supply chains | [13] | 2021 |
| Cloud robotics | Proposes a framework for DTws to control and improve precision | [30] | 2021 |
| Manufacturing | Uses ontologies to support multiscale modelling, with an industrial case study | [123] | 2021 |
| Industry 4.0 | Ontology modelling to support DTws for as-fabricated machining parts | [8] | 2021 |
| Surface roughness | Proposes human and machine readable dataset preparation method using semantic annotation | [107] | 2021 |
| Industry 4.0 | Focusses on requirements for DTws in industrial cyber-physical systems and proposes a Semantic Microservice Framework | [133] | 2021 |
| Smart factories | Proposes DTw-based method for an autonomous smart factory | [109] | 2021 |
| Ecosystem modelling | Proposes conceptual framework for decision support DTws to create value | [88] | 2021 |
| Manufacturing/I4.0 | Proposes DTw architecture for predicting control system runtime performance | [24] | 2021 |
| Industry 4.0 | Proposes conceptual framework for DTws to detect anomalies, with case study | [69] | 2021 |

| Context | Objective/Challenge resolved | Ref. | Year |
| --- | --- | --- | --- |
| virtual factory | Proposes a DTw-based approach to simulating and managing factories | [29] | 2021 |
| manufacturing | ontology-based modelling and automated DTw system planning and control | [52] | 2021 |
| Offsite Manufacturing | Compares approaches for Geometric DTws and proposes a combined solution | [83] | 2021 |
| Precision analysis | Proposes a method of analysing assembly precision through a Part DTw model | [134] | 2021 |
| Manufacturing | Develop architecture for HMI of autonomous industrial systems | [80] | 2021 |
| Assembly | Ontology-based method for modelling DTws for assembly processes | [37] | 2021 |
| manufacturing | Proposes a framework for DTws as and shows a case study | [17] | 2021 |
| Industrial Energy systems | Reviews DTw architectures and frameworks and shows proof of concept | [25] | 2020 |
| production systems | Proposes a framework for web-based DTws, with case study | [18] | 2020 |
| Industry 4.0 | Proposes a framework to integrate DTws with Manufacturing Execution Systems | [135] | 2020 |
| Smart Manufacturing | Framework for intelligent manufacturing DT | [38] | 2020 |
| Industry 4.0 | Looks to integrate DTws for optimisation/resilience in Factories of the Future | [44] | 2020 |
| Engineering systems | Proposes ontology-based DTw framework for connecting assets with test case | [53] | 2020 |
| I4.0 standards | Examine interoperability in existing DTw and IoT standards | [90] | 2020 |
| Industry 4.0 | Proposes a framework using DLTs and Privacy by Design | [1] | 2020 |
| Logistics | Survey defining DTws and exploring sources of value, validated through a case study in component logistics | [2] | 2020 |
| smart factories | Proposes a human-centric DTw and creates performance analysis testbeds | [19] | 2019 |
| Virtual rolling shop | Seeks to integrate tools to create ontology-based platform for digital factory | [26] | 2019 |
| process engineering | Proposes modular domain ontology, with a case study in control logic | [126] | 2019 |
| smart manufacturing | Demonstrator of architecture for open-source interoperable DTws | [9] | 2019 |
| Process safety | Applies systems thinking for how to improve DTw safety and interoperability | [14] | 2019 |
| Microstructure | Proposes context aware metho fog using a DTw for precise system control | [136] | 2019 |
| ***BUILT ENVIRONMENT*** | ***Architecture, Engineering, Construction, Operation-Facilities Management related articles*** | ***Σ=39*** | ***2020-23*** |
| AEC | Framework integrating BIM, XR, DTws, for construction progress monitoring | [78] | 2023 |
| HVAC | Proposes a DTw approach to fault detection, using AI to learn signs of failure | [110] | 2023 |
| Highways | Develops a semantic DTw risk management approach, with a case study | [137] | 2023 |
| Urban DTws | Systematic review and Delphi study of technical and non-technical issues | [51] | 2023 |
| Railways | Use ontologies to explore relationships between different railway features, with a case study for landslide failures | [85] | 2022 |
| Historic buildings | Applies cloud-based DTws and ontologies for conservation and maintenance | [21] | 2022 |
| Smart buildings | Proposes a framework to combine BIM and IoT using ontologies to access data | [106] | 2022 |
| Urban planning | Semantically fuses 3D point clouds and optical images to improve performance | [138] | 2022 |
| Prefabrication | Develops a framework for assessing quality between as-built and as-designed | [121] | 2022 |
| Fabrication | Systematic literature review identifying enablers, with a relational ontology to demonstrate linkages between enablers | [56] | 2022 |

| Context | Objective/Challenge resolved | Ref. | Year |
| --- | --- | --- | --- |
| Conservation | Develops systematic approach to building restoration, integrating BIM and DTws | [67] | 2022 |
| Urban streets | Uses DTws with point cloud segmentation segment different features | [139] | 2022 |
| Facility management | Combine BIM and ML for condition monitoring via sensors | [103] | 2022 |
| smart cities | Reviews how analytics could combine with DTws in smart city context | [10] | 2022 |
| Urban planning | Use pre-existing public data to create a geometric layer for city DTws | [105] | 2022 |
| AEC | Using real and synthetic datasets to integrate panoramic images into BIM | [129] | 2022 |
| smart city | Develop cloud system to integrate smart city data for incident management | [81] | 2022 |
| Urban Digital Twin | Creates a framework to evaluate and benchmark models of cities | [118] | 2022 |
| Urban geopolitics | Systematic literature review on how DTws interact with different tools | [63] | 2022 |
| District housing | Develops a workflow for creating DTws from GIS for energy assessment | [140] | 2022 |
| urban management | Proposes a method to improve accuracy of combining 2D with 3D in DTws | [54] | 2022 |
| Urban DTws | Literature review on how DTws and connected tools support urban sensing | [61] | 2022 |
| Smart cities | Literature review examining sustainable urban governance networks | [60] | 2022 |
| Smart Cities | Aggregates and anonymises ANPR data to be shared in DTws | [141] | 2022 |
| smart cities | Conducts bibliometric analysis to examine BIM-GIS integration for an urban DT | [108] | 2022 |
| Urban DTws | Systematic literature review of tools which support urban DTws | [62] | 2022 |
| Urban environments | Framework for city DTws to create synthetic datasets for training Deep Learning | [95] | 2022 |
| CDTws | Systematic literature review which proposes a CDTw reference architecture | [31] | 2021 |
| Built environment | Proposed user focussed DTw framework for dynamic sustainability assessment | [111] | 2021 |
| Buildings | Developed a method for DTws to use low-cost datasets for high quality models | [142] | 2021 |
| Lifecycle assessment | Literature review and conceptual framework for DTws to estimate embodied carbon across an asset lifecycle | [143] | 2021 |
| Tunnels | Proposed a maintenance DTw framework and a rule-based reasoning engine | [28] | 2021 |
| AECO-FM | systematic literature review exploring gaps and how DTws are currently used | [22] | 2021 |
| BLM | Investigated CDTws for BLM in the AEC sector | [79] | 2021 |
| Smart city | Developed a method to integrate BIM with GIS | [144] | 2021 |
| BIM | Framework for interoperable DTws integrating BIM for facility management | [76] | 2020 |
| smart cities | proposed architecture for semantic data processing and decision-making DTws | [75] | 2020 |
| construction | systematic literature review exploring the transition from BIM to DTws | [4] | 2020 |
| Asset management | integrating incomplete as-built data from BIM and IoT into DTws | [45] | 2020 |
| ***OTHER*** | ***Other sectors*** | ***Σ=18*** | ***2021-23*** |
| Metaverse | Reduce DTw energy consumption and integrate AI, Blockchain and XR | [77] | 2023 |
| Postal service | Runs scenarios with KPIs, in order to support optimisation | [145] | 2023 |
| Data management | Reference architecture for Findable, Accessible, Interoperable, Reusable DTws | [73] | 2023 |
| Energy | Proposes a DTw Platform for industrial energy systems | [40] | 2022 |
| Energy | Proposes an architecture to optimise energy efficiency via a reasoning engine | [113] | 2022 |

| Context | Objective/Challenge resolved | Ref. | Year |
|---|---|---|---|
| Healthcare | Proposes methods for Brain-Computer interfaces within DTws | [42] | 2022 |
| healthcare | Mathematically proposes a blockchain-based secure and private DTw model | [116] | 2022 |
| Smart devices | Proposes a DTw architecture for device provisioning and monitoring | [114] | 2022 |
| consumer IoT | Proposes decentralised architecture for DTws, ontologies and blockchain to support interoperability, security, resilience, and agility | [117] | 2022 |
| Cross-domain | Proposes framework for interoperability across chemical and electric domains | [125] | 2022 |
| Natural Disasters | Proposes a DTw framework using semantic analysis to better predict disasters | [128] | 2022 |
| libraries | Creates DTw model focussed on resource optimisation | [120] | 2022 |
| Aerospace | Proposes semantic framework for CDTw, with case study | [91] | 2022 |
| UAVs | Proposes a DTw model for intrusion detection systems | [87] | 2022 |
| Economics | Examine the economic implications of DTws and ontologies | [47] | 2021 |
| Aerospace | Literature review exploring what DTws are how they could apply to aerospace | [16] | 2021 |
| wind farms | Examines cyber-physical systems for safety, resilience, and reduced cost | [89] | 2021 |
| Virtual reality | Integrate VR into DTws and ontologies to improve robot monitoring and control | [72] | 2021 |
| ***GENERAL*** | ***Theoretical or no overarching domain*** | ***Σ=6*** | ***2021-22*** |
| General | Uses 4C architecture for DTws which integrate data across scale and scenario | [36] | 2022 |
| FAIR DTws | Proposes architecture for Findable, Accessible, Interoperable, Reusable DTws | [32] | 2022 |
| Maintenance | systematic lit review regarding CDTws with ontologies for maintenance | [6] | 2022 |
| Process modelling | Proposes an ontology framework to model objects across domains | [92] | 2021 |
| Digital twin web | Proposes open-source server to distribute and modify documents for DTws | [64] | 2021 |
| Systems Engineering | Develops a testbed using DTws to evaluate system performance | [119] | 2021 |